\documentclass[11pt]{article}
\usepackage{hyperref,amsmath,amsfonts,amssymb,eepic,graphicx,epsfig,epic}
\newcommand{\sepAuthor}{0.5in}
\newcommand{\sepAbstract}{0.4in}
\newcommand{\skipKeywords}{30pt}

\long\def\mytitlepage#1#2#3#4{
        \thispagestyle{empty}
        \begin{center}
        {\Large\bf #1}

        \vspace{\sepAuthor}
        #2\\
        \medskip

        \vspace{\sepAbstract}
        {\Large Abstract}
        \end{center}

        \noindent{#3}
        \vskip\skipKeywords

        \noindent{#4}
        \clearpage
        }
\usepackage{amsthm}
\theoremstyle{plain}
\newtheorem{theorem}{Theorem}[section]
\newtheorem{lemma}[theorem]{Lemma}
\newtheorem{corollary}[theorem]{Corollary}
\newtheorem{proposition}[theorem]{Proposition}

\theoremstyle{definition}
\newtheorem{definition}[theorem]{Definition}

\theoremstyle{remark}

\def\squareforqed{\hbox{\rlap{$\sqcap$}$\sqcup$}}
\def\qed{\ifmmode\squareforqed\else{\unskip\nobreak\hfil
\penalty50\hskip1em\null\nobreak\hfil\squareforqed
\parfillskip=0pt\finalhyphendemerits=0\endgraf}\fi}

\newenvironment{proofof}[1]{\begin{trivlist}%
\item[]{\flushleft\em Proof of #1. }}
{\qed\end{trivlist}}
\newcommand{\rank}{\textrm{rank}}
\newcommand{\delete}[1]{}
\newcommand{\LOCC}{\le_{\textrm{\tiny LOCC}}}
\newcommand{\SLOCC}{\le_{\textrm{\tiny SLOCC}}}
\def\>{\rangle}
\def\span{\textrm{span}}
\def\defeq{\stackrel{\textrm{\tiny def}}{=}}
\begin{document}
\mytitlepage{
When is there a multipartite maximum entangled state?
\footnote{This work was supported in part by the National Science
Foundation of the United States under Awards~0347078 and 0622033.
}}{Runyao Duan\footnote{
Email: \url{Runyao.Duan@uts.edu.au,dry@tsinghua.edu.cn}.
Supported in part by the National Natural Science Foundation of China
(Grant Nos. 60702080 and 60621062), the FANEDD under
Grant No. 200755, the Hi-Tech Research and Development Program
of China (863 project) (Grant No. 2006AA01Z102), and QCIS of UTS.
Part of this work was done while the author was visiting University of Michigan, Ann Arbor.
}\\
Centre for Quantum Computation and Intelligent Systems (QCIS),\\
Faculty of Engineering and Information Technology,\\
University of Technology, Sydney (UTS), NSW 2007, Australia \\
and\\
Department of Computer Science and Technology,\\
Tsinghua National Laboratory for Information Science and
Technology, \\
Tsinghua University, Beijing 100084, China\\
\vspace{2ex}

Yaoyun Shi\footnote{Email: \url{shiyy@eecs.umich.edu}.}\\
Department of Electrical Engineering and Computer Science,
University of Michigan\\
2260 Hayward Street, Ann Arbor, MI 48109-2121, USA
}{\noindent For a multipartite quantum system of the dimension $d_1\otimes d_2\otimes\cdots d_n$,
$d_1\ge d_2\ge\cdots\ge d_n$, is there an entangled state {\em maximum} in the sense
that all other states in the system can be obtained from the state through local
quantum operations and classical communications (LOCC)?
When $d_1\ge\Pi_{i=2}^n d_i$, such state exists. We show that this condition is also necessary.
Our proof, somewhat surprisingly, uses results from algebraic complexity theory.}

\newcommand{\EPR}{|\Psi^-\rangle}
\newcommand{\bra}[1]{\langle #1|}
\newcommand{\ket}[1]{|#1\rangle}
\newcommand{\braket}[3]{\langle #1|#2|#3\rangle}
\newcommand{\ip}[2]{\langle #1|#2\rangle}
\newcommand{\op}[2]{|#1\rangle \langle #2|}

\newcommand{\tr}{{\rm tr}}
\newcommand{\supp}{{\it supp}}
\newcommand{\sch}{{\it Sch}}

\newcommand{\Span}{\mathrm{span}}
\newcommand {\E } {{\mathcal{E}}}
\newcommand{\In}{\mathrm{in}}
\newcommand{\Out}{\mathrm{out}}
\newcommand{\local}{\mathrm{local}}
\newcommand {\F } {{\mathcal{F}}}
\newcommand {\diag } {{\rm diag}}
\renewcommand{\b}{\mathcal{B}}
\newcommand{\h}{\mathcal{H}}
\renewcommand{\Re}{\mathrm{Re}}
\renewcommand{\Im}{\mathrm{Im}}
\newcommand{\Z}{\sigma_z}
\newcommand{\Clocal}{C^{(0)}_{\local}}
\newcommand{\Sp}[1][p]{S_{\min}^{(#1)}}

\section{Background and the statement of the main result}
A quantum system consisting of several subsystems may be in an {\em entangled}
state, such that measurements on the subsystems may produce outcome
statistics fundamentally different from those produced through a classical process. Since its discovery by Einstein, Podolsky, and
Rosen~\cite{EPR35},
quantum entanglement has been found to be central for non-classical properties
of quantum systems. In particular, it plays a fundamental role in quantum information processing applications
such as unconditional secure key distribution and super fast quantum algorithms.
It is therefore of fundamental importance to understand the nature of entanglement.
Indeed, the past two decades have witnessed the rapid development of a theory of quantum information,
at the heart of which is the theory of quantum entanglement. Horodecki {\em et al.} \cite{Horodecki}
and G{\"u}hne and T{\'o}th~\cite{Guhne} are recent surveys on the subject.

Our study is motivated by the following objective, which is important for the practical applications
of quantum entanglement: how do we establish quantum entanglement
between multiple parties separated spatially? One straightforward solution is for one party to prepare the desired
state $|\phi\>$, and send the others their corresponding portion of the state.
The problem of this solution is that moving quantum objects around without
corrupting them is difficult and expensive, especially when the parties are remotely
separated.

The celebrated {\em quantum teleportation protocol} \cite{BBCJ+93}
provides an alternative approach: the parties initially share some special but fixed
entangled state $|\phi_0\>$, which will then be transformed to $|\phi\>$ through
local quantum operations and classical communications (LOCC). Ideally, $|\phi_0\>$
should work for all possible $|\phi\>$ desired.
The question we address is: for which dimensions of the system is there such an initial
state that can generate all other states in the system?

Let $n, d_1, d_2, \cdots, d_n$ be integers with $n\ge2$,
and $d_1\ge d_2\ge \cdots \ge d_n\ge 2$. We denote by $d_1\otimes d_2\otimes\cdots
d_n$ the tensor product of $n$ Hilbert spaces, each of the dimension $d_1$, $d_2$, ..., $d_n$,
respectively. We refer to the whole system by $H$, and each subsystem by $A$, $B$, $C$, ..., $Z$,
respectively.
We use superscripts $A, B, C, \cdots, Z$ on states or operators to indicate the space
they are associated with.
Let $|\phi_1\>$ and $|\phi_2\>$ be two states in $d_1\otimes\cdots\otimes d_n$.
We write $|\phi_2\>\LOCC|\phi_1\>$ if $|\phi_1\>$ can be transformed
to $|\phi_2\>$ through a LOCC protocol. A state $|\phi_0\>$ is said to be a
maximum entangled state (MES)
if $|\phi\>\LOCC|\phi_0\>$ for all $|\phi\>$ in the same space.
Thus our problem is, which space $d_1\otimes\cdots\otimes d_n$
contains a maximum entangled state?

Besides the practical motivation described above, our question
is also among the most basic questions in the framework
of entanglement manipulations, which is to study properties
of entanglement under LOCC transformations. This is a major paradigm
for studying entanglement where many central results were obtained.
A particular task in this paradigm is to classify entangled states through their conversion
relations. Note that classical communication
should not increase any reasonable notion quantifying entanglement --- indeed,
this monotonicity under LOCC transformation is considered
the only natural requirement for entanglement measures~\cite{Vidal}.
Therefore, the relation $\LOCC$ induces a natural partial ordering of quantum
states (or more precisely, of the LOCC equivalence classes) by the amount of entanglement. Our problem, which is to ask when a maximum
element exists, is thus among the very basic questions regarding the structure of this ordering.
We stress that the definition of ``maximum'' in this paper is restricted to the LOCC
ordering. There may be other definitions of maximum entangled states with respect
to other orderings.

When $n=2$, the answer to our question is well known through the use of the teleportation protocol
with the generalized EPR state (commonly referred to as the maximum entangled state
for bipartite systems)
\begin{equation}\label{eqn:EPR}
|\Phi_{d_2}\>^{AB}\defeq \sum_{i=0}^{d_2-1}|i\>^A|i\>^B,
\end{equation}
where $\{|i\>^A: i=0..d_2-1\}$ and $\{|i\>^B: i=0..d_2-1\}$ are
orthonormal in $A$ and $B$, respectively.
The teleportation protocol can be generalized to arbitrary $n$, as long as
\begin{equation}\label{eqn:main}
d_1\ge \Pi_{i=2}^K d_i.
\end{equation}

On the other hand, not all spaces have a maximum entangled state.
For example, D{\"u}r, Vidal and Cirac showed that there is no MES
in the $2\otimes2\otimes2$ space \cite{DVC00}.
The main result of this paper is that
a MES exists only if Eqn.~(\ref{eqn:main}) holds.

Our result is actually slightly stronger. Following the notation of
Bennett {\em et al.}~\cite{BPRS+00}, if $|\phi_1\rangle$
can be transformed to $|\phi_2\>$ with a non-zero probability, we
write $|\phi_2\>\SLOCC|\phi_1\>$, where ``SLOCC'' stands for
{\em Stochastic} Local Operations and Classical Communications.
Similarly, $|\phi_1\>$ is called a stochastic maximum entangled state
if $|\phi_2\>\SLOCC|\phi_1\>$ for all $|\phi_2\>$ in the same space.
The partial ordering $\SLOCC$ was introduced by Bennett et al. \cite{BPRS+00}
in order to provide a simpler classification of multipartite entanglement
(there are infinitely number of LOCC equivalence classes even for $2$ qubits),
and has been subsequently studied by many authors.
Clearly a MES is also a SMES; thus if Eqn.~(\ref{eqn:main}) holds
then a SMES exists. We now state our main theorem.

\begin{theorem}\label{thm:main} If $d_1<\Pi_{i=2}^n d_i$,
there is no stochastic maximum entangled state in the state space $d_1\otimes d_2\otimes\cdots\otimes
d_n$.
\end{theorem}

Our proof uses the notion of tensor rank
from algebraic complexity theory (C.f. Chapter 14 in \cite{Burgisser}).
The tensor rank of $\ket{\phi}\in \h$,
$\sch(\phi)$, is the minimum number of product vectors that can
linearly express $\ket{\phi}$. That is, $\sch(\phi)$ is the minimum
integer $k$ such that there exists product vectors
$|\phi^A_i\>\otimes|\phi^B_i\>\otimes\cdots|\phi^Z_i\>\in d_1\otimes d_2\otimes\cdots\otimes d_n$
such that
$$\ket{\phi}\defeq\sum_{i=1}^k|\phi^A_i\>\otimes|\phi^B_i\>\otimes\cdots|\phi^Z_i\>.$$
$\sch(\phi)$ can also be called {\em Schmidt rank}
or {\em Schmidt number} \cite{EB00}, and is precisely the rank of the
reduced density matrix $Tr_{A}(\op{\phi}{\phi})$ when $n=2$.
In general, the tensor rank is the minimum number
of multiplications to compute a set of linear forms determined by
$\ket{\phi}$. For example, the minimum number of non-scalar multiplications
for multiplying two $n$ by $n$ matrices is precisely the tensor rank
of the following element in $n^2\otimes n^2\otimes n^2$:
$$\sum_{i,j, k=0}^{n-1}|i, j\>|i, k\>|k, j\>,$$
where each component space has a product orthonormal basis $\{|i,j\>: i, j=0..n-1\}$.
It was observed in \cite{CDS08} that the above state is precisely the tripartite state $\ket{\Psi_n}^{ABC}=\ket{\Phi_n}^{AB}\otimes\ket{\Phi_n}^{BC}\otimes \ket{\Phi_n}^{CA}$. This connection enables us and a co-author to show the equivalence
between the computational complexity of matrix multiplication and efficiency of
a certain entanglement transformation that produces EPR pairs~\cite{CDS08}.

The tensor rank of a Hilbert space $\h$ is
$$\sch(\h)\defeq\max\{\sch(\phi):\ket{\phi}\in \h\}.$$
Many works have been done to determine the tensor rank of specific tensors
and of various spaces. We will use the following results.

\begin{theorem}\label{thm:mtr} Consider $\sch(\h)$ for $\h=d_1\otimes d_2\otimes d_3$.
Let $k=d_2d_3-d_1$.
\begin{itemize}
\item[(i)] (Theorem 6(ii) of \cite{GAT91}) If $k\geq 1$, then $\sch(\h)\geq d_1+\lfloor
\sqrt{2k+2}\rfloor-2$.
\item[(ii)] (Theorem 3 of \cite{BSH}) If $k\leq \max\{d_2,d_3\}$ and $0\leq k\leq 4$,  then
$\sch(\h)=d_2d_3-\lceil\frac{k}{2}\rceil$.
\end{itemize}
\end{theorem}

\section{Proof of the Main Theorem}
We now turn to the proof of the main result. We shall first obtain
some structural results about SLOCC and the induced ordering on the states.
We say that $|\phi_1\>$ and $|\phi_2\>$ are SLOCC equivalent
if $|\phi_1\>\SLOCC|\phi_2\>$ and $|\phi_2\>\SLOCC|\phi_1\>$.
Then $\SLOCC$ defines a partial oder on SLOCC equivalence
classes. We will often identify a state with its equivalence class.
A state $|\phi\>$ is said to be SLOCC {\it maximal} if for any $|\psi\>$,
$|\phi\>\SLOCC|\psi\>$ implies $|\psi\>\SLOCC|\phi\>$. For the rest
of the paper, we may omit ``SLOCC'' when referring to equivalence,
equivalence classes, maximal state, etc.
We know the following fact about SLOCC \cite{DVC00}.

\begin{lemma}\label{lemma1}\upshape
Let $\ket{\phi}$ and $\ket{\psi}\in d_1\otimes d_2\cdots \otimes d_n$.
Then $\ket{\psi}\SLOCC\ket{\phi}$ if and
only if there are linear operators $L_1,\cdots, L_n$ such that
$(L_1\otimes\cdots \otimes L_n)\ket{\phi}=\ket{\psi}$. In
particular, $\ket{\phi}$ and $\ket{\psi}$ are equivalent under SLOCC
if and only if $L_1,\cdots, L_n$ are invertible.
\end{lemma}

Since local linear operators cannot increase tensor rank,
we have the following fact that relates tensor rank and SLOCC \cite{LP97}.
\begin{proposition}\label{prop:tr}
If $\ket{\psi}\SLOCC\ket{\phi}$, $\sch(\phi)\geq \sch(\psi)$.
\end{proposition}

We say that $\ket{\Phi}\in \h$ is of full local ranks if
${\rank}(\rho^{k}_{\Phi})=d_k$ for any $k$, $1\leq k\leq n$, where
$\rho^{k}_{\Phi}$ is the reduced density operator of $\op{\Phi}{\Phi}$
obtained by tracing out all subsystems other than the $k$'th one.
We characterize maximal states below.
\begin{lemma} A state is maximal if and only if it has full local ranks.
\end{lemma}

\begin{proof} We prove the result for $n=3$. The other cases are similar.
Suppose that $\ket{\Phi}$ is of full local ranks. Let
$\ket{\Psi}\in \h$  be such that $\ket{\Phi}\SLOCC\ket{\Psi}$.
Then there exists linear
operators $L_1,L_2,L_3$ such that $(L_1\otimes L_2\otimes
L_3)\ket{\Psi}=\ket{\Phi}$.  As $\ket{\Phi}$ is of full local ranks,
we have that $L_1,L_2,L_3$ should be invertible. Thus
$\ket{\Psi}$ and $\ket{\Phi}$ are equivalent, implying
that $\ket{\Phi}$ is maximal.

For the other direction, assume for the purpose of getting a contradiction
that $\ket{\Phi}$ is maximal but is not of full local ranks.
 Without loss of generality , assume that
$\rank(\rho_{\Phi}^{1})=k<d_1$. Let $\{\ket{i}^{A}:1\leq i\leq
k\}$ be an orthonormal basis for the support of $\rho_{\Phi}^{A}$.
Write
$$\ket{\Phi}=\sum_{i=1}^{k}\ket{i}^{A}\ket{\phi_i}^{BC}.$$
 Let
$\ket{\psi}^{A}\in A$ be such that $\ip{i}{\psi}=0$ for all $i$, $1\leq
i\leq k$.  Construct $\ket{\Phi'}$ as follows:
$$\ket{\Phi'}=\ket{\Phi}+\ket{\psi}^{A}\ket{\phi}^{BC},$$
where $\ket{\phi}^{BC}$ is any nonzero vector. Let
$P^{A}=\sum_{i=1}^{k}\op{i}{i}$. One can easily
verify that $\ket{\Phi'}$ can be transformed into $\ket{\Phi}$ by
SLOCC as $$(P^{A}\otimes I^{B}\otimes
I^{C})\ket{\Phi'}=\ket{\Phi}.$$ However, $\ket{\Phi'}$ is not
equivalent to $\ket{\Phi}$  as
$\rank(\rho_{\Phi'}^{A})=k+1>\rank(\rho_{\Phi}^{A})$. That
contradicts the fact that $\ket{\Phi}$ is maximal.
\end{proof}

The partition of the space $d_1\otimes\cdots\otimes d_n$ into $n$ sub-systems
may be further refined by partitioning one, or several, sub-system $d_i$ into
a product space $d_{i, 1}\otimes d_{i, 2}\otimes\cdots \otimes d_{i, k_i}$,
where $\Pi_{j=1}^{k_i} d_{i, j}=d_i$. Note that if a density operator
$\rho^i$ on the $i$'th sub-system is of full rank, its reduced density
operator $\rho^{i, j}$, $j=1..k_j$ on the $j$'th sub-system in the refinement
is also of full rank. Thus we have the following
useful consequence of the above lemma.
\begin{corollary}\label{co:refine}
A maximal state remains maximal with respect to a refined partition.
\end{corollary}

The following lemma shows that there are at least
two general ways of constructing a maximal state.
\begin{lemma}\label{twomax}\upshape
There is a maximal state $\ket{\Phi}$ such that
$\sch(\Phi)=\sch(\h)$. There is also a maximal state with tensor
rank $d_1$.
\end{lemma}

\begin{proof}
By definition, there exists $\ket{\Phi}$ such that
$\sch(\Phi)=\sch(\h)$. So we can write
$$\ket{\Phi}=\sum_{i=1}^{\sch(\h)}\ket{\alpha_i}\ket{\beta_i}\ket{\gamma_i}.$$
If $\ket{\Phi}$ is of full local rank, then it follows from the
previous lemma that $\ket{\Phi}$ is maximal. Otherwise, assume
without loss of generality that $\rank(\rho_\Phi^{A})<d_1$. Thus
there exists another nonzero vector $\ket{a_1}\in \h_{A}$ such that
$\ip{a_1}{\alpha_i}=0$, for all $i$. Construct $\ket{\Phi'}$ as follows:
$$\ket{\Phi'}=\ket{\Phi}+\ket{a_1}\ket{b_1}\ket{c_1},$$
where $\ket{b_1}\in \h_{B}$ and $\ket{c_1}\in \h_{C}$ are arbitrary
non-zero vectors. Note that we can obtain $\ket{\Phi}$ from $\ket{\Phi'}$ by
performing the local projection $(I-\op{a_1}{a_1})^{A}\otimes
I^{B}\otimes I^{C}$. Thus $\sch(\Phi')\geq \sch(\Phi)$, and that
$\ket{\Phi'}$ is of the maximal tensor rank $\sch(\h)$. Furthermore
we have $\rank(\rho_{\Phi'}^{A})>\rank(\rho_{\Phi}^{A})$. If
$\ket{\Phi'}$ is of full local rank already, the proof is complete.
Otherwise repeat the above arguments.
Thus after a finite number of repetitions of the above steps,
we can obtain a state of the maximum tensor rank
and, of full local ranks, thus maximal.

We construct a maximal state with the tensor rank $d_1$ as follows.
Take a basis $\{\ket{a_k}:k=1,\cdots, d_1\}$ of $\h_{A}$ and a set of
$d_1$ linearly independent product vectors
$\{\ket{b_k}\ket{c_k}:1\leq k\leq d_1\}$ of $\h_{B}\otimes
\h_{C}$ and then construct
$$\ket{\Psi}=\sum_{k=1}^{d_1}\ket{a_k}\ket{b_k}\ket{c_k}.$$
Clearly $\sch(\Psi)=d_1$. However, we cannot guarantee that
$\sch(\Psi)$ is  of full local rank at $B$ and $C$ sides. For
instance,  $\{\ket{b_k}:1\leq k\leq d_1\}$ may not span $\h_{B}$.
A simple example is $\ket{0}\ket{00}+\ket{1}\ket{01}$, which is of
tensor rank $2$ but is not of full rank.  One can avoid this problem by
using the special construction presented in Ref. \cite{DFJY07}. An alternative
construction is as follows. Let $\{\ket{0}, \cdots, \ket{d-1}\}$ be a basis for a dimension $d$ space.
Consider the state
\[\sum_{i=0}^{d_3-1} \ket{i, i, i} + \sum_{i=d_3}^{d_2-1} \ket{i, i, 0} +
\sum_{i=d_2}^{d_1-1}\ket{i, a_i, c_i},\]
where $(a_i, c_i)$'s are distinct elements that do not appear in the first two terms.
It is quite straightforward to verify the above state is maximal and has tensor rank $d_1$.
\end{proof}

We will assume from now on that $d_1<d_2\cdots d_n$, and show
that there are at least two incomparable maximal states under
this assumption. We will focus on $n=3$ and return to the general case later.
Let $k=d_2d_3-d_1$.

First, we prove the result for the case $\sch(\h)>d_1$.
We then show if $d_1<d_2d_3-1$ then $\sch(\h)>d_1$. Finally
we show that for $d_1=d_2d_3-1$, there are precisely $\min\{d_2, d_3\}\ge2$
number of maximal equivalence classes.

By Lemma~\ref{twomax}, if $\sch(\h)\ne d_1$ then there
are two incomparable maximal states.
This is indeed the case when $k>1$.
\begin{lemma}\label{lm:smallk}
There are at least two incomparable maximal states
in $d_1\otimes d_2\otimes d_3$ if $k>1$.
\end{lemma}

\begin{proof} By Theorem~\ref{thm:mtr}, we have $\sch(\h)\geq d_1+1$ for $k\ge 4$, by Item (i),
and for $k=2, 3$ by Item (ii) (note that when $k=3$,
$\max\{d_2, d_3\}\ge k$, since otherwise $d_1=d_2=d_3=k=2$).
Therefore, when $k>4$, $\sch(\h)\ne d_1$.
Since any two equivalent states must have the same tensor rank
(by Proposition~\ref{prop:tr}), Theorem~\ref{thm:main} implies that there
are two incomparable maximal states.
\end{proof}

We now focus on the case $d_1=d_2d_3-1$.
By Theorem~\ref{thm:mtr}(ii), $\sch(\Phi)=d_1$. Since
a maximal state has full local ranks thus having a tensor
rank $\ge d_1$, its tensor rank must be precisely $d_1$.
The following lemma completes the proof for our main
theorem for $n=3$.

\begin{lemma}\label{bipartite}\upshape
If  $d_1=d_2d_3-1$, there are precisely $\min\{d_2,d_3\}$ inequivalent maximal
states.
\end{lemma}

\begin{proof}
Let $\ket{\Phi}=\sum_{i=0}^{d_1-1}\ket{i}^{A}\ket{\phi_i}^{BC}$ and
$\ket{\Psi}=\sum_{i=0}^{d_1-1}\ket{i}^{A}\ket{\psi_i}^{BC}$
are two maximal states. Since they are of full local ranks,
$\{|\phi_i\>:i=0..d_1-1\}$ and $\{|\psi_i\>:i=0..d_1-1\}$
are linear independent. Therefore, there exist two unique
(up to a phase factor) non-zero states $\ket{\Phi'}^{BC}$ and
$\ket{\Psi'}^{BC}$ such that $\ip{\Phi'}{\phi_i}=0$ and
$\ip{\Psi'}{\psi_i}=0$, for all $i=0..d_1-1$.

If $\ket{\Phi}$ and $\ket{\Psi}$ are equivalent,
there exist invertible operators $L_1,L_2,L_3$ such
that $\ket{\Phi}=(L_1\otimes L_2\otimes L_3)\ket{\Psi}$,
implying that
\begin{equation}\label{eqn:complement}
\span\{\ket{\phi_i}:1\leq i\leq d_1\}=(L_2\otimes L_3)\span\{\ket{\psi_i}:1\leq i\leq d_1\}.
\end{equation}
This is equivalent to
\begin{equation}\label{eqn:complement_state}
\ket{\Phi'}^{BC}=((L_2^\dagger)^{-1}\otimes
(L_3^\dagger)^{-1})\ket{\Psi'}^{BC}.
\end{equation}
In other words,
$\ket{\Phi'}$ and $\ket{\Psi'}$ are equivalent under SLOCC. Note
that $\ket{\Phi'}$ and $\ket{\Psi'}$ are both bipartite pure states.
It is well known that two bipartite pure states are SLOCC equivalent
if and only if they have the same Schmidt rank.
For a non-zero pure state in $d_2\otimes d_3$, the Schmidt rank
may take any values from $1$, ..., $\min\{d_2, d_3\}$.
Therefore, there are at least
$\min\{d_2, d_3\}$ different equivalence classes of stochastic
maximal states.

On the other hand, if $\ket{\Phi'}$ and $\ket{\Psi'}$ have
the same Schmidt rank, there exist invertible $L_2$ and $L_3$ such
that Eqn.~(\ref{eqn:complement_state}), and consequently, Eqn.~(\ref{eqn:complement})
hold. Thus $L_2\otimes L_3\ket{\Psi}=\sum_{i=0}^{d_1-1}\ket{\alpha_i}^{A}|\psi_i\rangle$,
for some states $\ket{\alpha_i}^A$, $i=0..d_1-1$. Those states must
be linearly independent, since $L_2\otimes L_3$ does not change the local rank of
$\ket{\Psi}$. Thus setting $L_1=\sum_{i=0}^{d_1-1}|i\rangle\langle\alpha_i|$, we have that
$L_1$ is invertible and $\ket{\Phi}=L_1\otimes L_2\otimes L_3 \ket{\Psi}$.
Thus $\ket{\Phi}$ and $\ket{\Psi}$ are equivalent. Consequently,
there are at most $\min\{d_2, d_3\}$ number of maximal equivalence class.
\end{proof}

An example to illustrate Lemma \ref{bipartite} is the
state space $\h=\h_3\otimes \h_2\otimes \h_2$. Miyake
has obtained all eight equivalence class of this space \cite{MIY03}.
Two of these equivalence classes are maximal. The above Lemma provides an alternative
method to characterize the maximal states in this space. By the Lemma,
there is a one-to-one correspondence between the
maximal equivalence class of $\h$ and the equivalence class of
$\h'=\h_2\otimes\h_2$. The latter space has precisely two equivalence class
with the representatives $\ket{\Phi_1'}=\ket{10}$ and
$\ket{\Phi_2'}=\ket{01}-\ket{10}$. As a result, there are only two
maximal equivalence class, which can be constructed according to
$\ket{\Phi_1'}$ and $\ket{\Phi_2'}$ as follows:
\begin{eqnarray}
\ket{\Phi_1}&=&\ket{0}\ket{00}+\ket{1}\ket{01}+\ket{2}\ket{11},\nonumber\\
\ket{\Phi_2}&=&\ket{0}\ket{00}+\ket{1}(\ket{01}+\ket{10})+\ket{2}\ket{11}.\label{eqn:states223}
\end{eqnarray}

Lemma~\ref{lm:smallk} and \ref{bipartite} together imply Theorem~\ref{thm:main}.

We have finished the proof of Main Theorem for $n=3$. We deal with the general
case below (that is to show that there is no maximum state in $d_1\otimes d_2\otimes\cdots\otimes d_n$
if $d_1<d_2d_3\cdots d_n$).

\begin{proofof}{Theorem\ref{thm:main}} We need only consider
$n>3$. Suppose that $n=4$ and $d_1<d_2d_3d_4$. Consider the
tripartite state space $d_1\otimes d_2\otimes
d_3d_4$. There are two cases:

Case 1. $d_3d_4=d_1d_2$. Since $d_1\geq d_2\geq d_3\geq d_4$ we have
$d_1=d_2=d_3=d_4=d$. One can easily verify that
$\ket{\Phi_d}^{AB}\otimes \ket{\Phi_d}^{CD}$ and
$\ket{\Phi_d}^{AC}\otimes \ket{\Phi_d}^{BD}$ both are of
full local rank $d$, where $\ket{\Phi_d}$ is the generalized
EPR state defined in Eqn.~\ref{eqn:EPR}.
Observe that with respect to the $AC:BD$ partition, the former is entangled
yet the latter is not, and with respect to the $AB:CD$ partition,
the opposite holds. Since no LOCC protocol can create entanglement,
the two states are not comparable under SLOCC.

Case 2. $d_3d_4<d_1d_2$. In this case, applying the result for $n=3$
we know that there are at least two inequivalent maximal states in
${d_1}\otimes {d_2}\otimes {d_3d_4}$. By Corollary~\ref{co:refine},
those states are also maximal in $d_1\otimes d_2\otimes d_3\otimes d_4$.
They remain incomparable under SLOCC with respect to this refined partition.

Suppose that the theorem is correct for $n=k$, $k\ge4$. Consider $n=k+1$.
Since $k\ge4$, $d_kd_{k+1}<d_1d_2d_3\cdots d_{k-1}$. By the inductive hypothesis,
there are two incomparable maximum states in $d_1\otimes d_2\otimes\cdots \otimes d_{k-1}
\otimes d_kd_{k+1}$. By Corollary~\ref{co:refine}, they remain maximal
and incomparable in the refinement $\h$. Thus the theorem is correct for $n=k+1$,
therefore correct for all $n\ge4$.
\end{proofof}

\section{Correspondence between maximal equivalence class and SLOCC  equivalences classes}

In this section we consider state spaces such that
$d_1<d_2\cdots d_K$. So it is impossible
to find one state from which one can locally prepare any other state even
probabilistically.

An alternative goal is to characterize
all maximal equivalence class. In particular, we ask when
a multipartite state space $\h$ has only a finite number of maximal stochastic equivalence classes.
Suppose that $\h$ has a finite number of
maximal equivalence class with the representative states
$\ket{\Phi_1},\cdots, \ket{\Phi_N}$. Then for any state
$\ket{\psi}\in \h$, there exists $1\leq k\leq N$ such that
$\ket{\Phi_k}$ can be converted into $\ket{\psi}$ by SLOCC. So the
set of states $\{\ket{\Phi_1},\cdots, \ket{\Phi_N}\}$ is able to
locally prepare any other state in $\h$ with nonzero probability. In
practice, we only need to prepare the set of maximal states
$\{\ket{\Phi_k}\}$ and then create other states using SLOCC. Thus
identifying the maximal equivalence classes for a given space is highly
desirable.

For the sake of convenience, from now on we mainly focus on
tripartite state space. Most of our results are also valid for the
case of $K>3$. We assume that $d_1=d_2d_3-k$, where $k<d_2d_3/2$. We
shall employ a correspondence between the maximal equivalence class of
$\h_{d_1}\otimes \h_{d_2}\otimes \h_{d_3}$ and the equivalence class of $\h_{k}\otimes \h_{d_2}\otimes \h_{d_3}$.

\begin{definition}\upshape
Let $\ket{\Phi}\in \h_{d_1}\otimes \h_{d_2}\otimes\h_{d_3}$ and
$\rank(\rho_{\Phi}^{A_1})=d_1$, write
$\ket{\Phi}=\sum_{i=1}^{d_1}\ket{i}^{A_1}\ket{\phi_i}^{A_2A_3}$,
where $\{\ket{i}^{A_1}:1\leq i\leq d_1\}$ is any basis for
$\h_{d_1}$. Let $\mathcal{T}^{A_1}(\Phi)$ be the SLOCC
equivalence class of $\h_{k}\otimes \h_{d_2}\otimes\h_{d_3}$ with
representative state
$\ket{\Phi'}=\sum_{i=1}^{k}\ket{i}^{A_1}\ket{\phi_i^{\perp}}^{A_2A_3}$,
where $\{\ket{i}^{A_1}:1\leq i\leq k\}$ is a basis for $\h_k$ and
$\{\ket{\phi_i^{\perp}}:1\leq i\leq k\}$ is any basis for ${\rm
span}^{\perp}\{\ket{\phi_i}^{A_2A_3}:1\leq i\leq d_1\}$.
\end{definition}
It is easy to see that $\mathcal{T}^{A_1}({\Phi})$ is
well-defined in the sense that it does not depend on which basis of
${\span}^{\perp}\{\ket{\phi_i}^{A_2A_3}:1\leq i\leq d_1\}$ we
choose. It is also worth noting that any state in $\mathcal{T}^{A_1}({\Phi})$ should be of
local rank $k$ between $A_1$ and $A_2A_3$.

The importance of the map $\mathcal{T}^{A_1}$ is due to the
following lemma, which can be treated as a generalization of Lemma
\ref{bipartite}.
\begin{lemma}\label{key}\upshape
Let $\ket{\Phi}$ and $\ket{\Psi}$ be two vectors in $\h$ such that
$\rank(\rho_\Phi^{A_1})=\rank(\rho_\Psi^{A_1})=d_1$. Then $\ket{\Phi}$
and $\ket{\Psi}$ are equivalent under SLOCC if and only if
$\mathcal{T}^{A_1}({\Phi})=\mathcal{T}^{A_1}({\Psi})$.
\end{lemma}

\begin{proof} The proof idea is similar to Lemma \ref{bipartite}. For
completeness, we present a detailed proof here. By Lemma
\ref{lemma1}, $\ket{\Phi}$ and $\ket{\Psi}$ are equivalent under
SLOCC if and only if there are invertible linear operators $L_1,
L_2,L_3$ such that $\ket{\Phi}=(L_1\otimes L_2\otimes
L_3)\ket{\Psi}$. More explicitly, we have
\begin{equation}
\sum_{i=1}^{d_1}\ket{i}^{A_1}\ket{\phi_i}^{A_2A_3}=(L_1\otimes
L_2\otimes L_3)\sum_{i=1}^{d_1}\ket{i}^{A_1}\ket{\psi_i}^{A_2A_3}.
\end{equation}
Applying $\bra{j}^{A_1}\otimes I^{A_2A_3}$ to both sides of the
above equation, we have
$$\ket{\phi_j}^{A_2A_3}=(L_2\otimes L_3)\sum_{i=1}^{d_1}\braket{j}{L_1}{i}\ket{\psi_i}^{A_2A_3}.$$
That means $$\ket{\phi_j}^{A_2A_3}\in (L_2\otimes
L_3)\span\{\ket{\psi_i}^{A_2A_3}:1\leq i\leq d_1\}$$ for each $1\leq
j\leq d_1$. Noticing further that $L_1$ is invertible, we have
\begin{equation}\label{sloccsubspace}
\span\{\ket{\phi_i}^{A_2A_3}\}=(L_2\otimes
L_3)\span\{\ket{\psi_i}^{A_2A_3}\}.
\end{equation}
Conversely, we can readily show that the existence of invertible
linear operators $L_2$ and $L_3$ such that Eqn. (\ref{sloccsubspace})
holds also implies the SLOCC equivalence between $\ket{\Phi}$ and
$\ket{\Psi}$. It is easy to verify Eqn. (\ref{sloccsubspace}) can be
rewritten into the following
\begin{equation}\label{sloccsubspace2}
\span^\perp\{\ket{\phi_i}^{A_2A_3}\}=((L_2^{\dagger})^{-1}\otimes
(L_3^{\dagger})^{-1})\span^\perp\{\ket{\psi_i}^{A_2A_3}\}.
\end{equation}
Using a similar argument, we can show the above equation means that
$\ket{\Phi'}=\sum_{i=1}^{k}\ket{i}^{A_1}\ket{\phi_i^{\perp}}^{A_2A_3}$
and
$\ket{\Psi'}=\sum_{i=1}^{k}\ket{i}^{A_1}\ket{\psi_i^{\perp}}^{A_2A_3}$
are equivalent. In other words,
$\mathcal{T}^{A_1}({\Phi})$ and $\mathcal{T}^{A_1}({\Psi})$
coincide.
\end{proof}

When $k<d_2d_3/2$,  we have $k<d_1$. It may be much
easier to
decide the SLOCC equivalence between $\mathcal{T}^{A_1}({\Phi})$
and $\mathcal{T}^{A_1}({\Psi})$ than that between $\ket{\Phi}$
and $\ket{\Psi}$. However, $\mathcal{T}^{A_1}$ is not a one-to-one
correspondence between the maximal equivalence class
of $\h_{d_1}\otimes \h_{d_2}\otimes\h_{d_3}$ and the
equivalence class of
$\h_k\otimes\h_{d_2}\otimes\h_{d_3}$. In general, the image of
$\mathcal{T}^{A_1}$ is only a proper subset of $\h_{k}\otimes
\h_{d_2}\otimes\h_{d_3}$. Fortunately, in the special case of $k=1$,
we do have a one-to-one correspondence as stated below.
The case of $n=3$ was proved in
in Lemma \ref{bipartite}.
\begin{theorem}\label{one-to-one}\upshape
There is a one-to-one correspondence between the maximal equivalence class
in $\h_{d_1}\otimes \cdots\otimes
\h_{d_K}$ and the stochastic equivalence classes of $\h_{d_2}\otimes
\cdots\otimes \h_{d_K}$, where $d_1=d_2\cdots d_K-1$.
\end{theorem}

The following theorem also follows directly from Lemma \ref{key}.
\begin{theorem}\label{finite}\upshape
If $\h_k\otimes \h_{d_2}\otimes\cdots\otimes \h_{d_K}$ has a finite
number of equivalence class, then $\h_{d_1}\otimes
\h_{d_2}\otimes\cdots\otimes \h_{d_K}$ has a finite number of
maximal equivalence class.
\end{theorem}

Using the known result that there are a finite number
of equivalence class for tripartite systems of the
dimensions $d_3=2$, $d_2\le 3$~\cite{Chen}, we have the following corollary.

\begin{corollary}\label{special.finite}\upshape
Each of the following spaces has a finite number of maximal equivalence class:
$(2n-2)\otimes n\otimes 2$, $(2n-3)\otimes n\otimes 2$, $(3n-2)\otimes n\otimes 2$,
and, when $2\le \min\{m, n\} \le 3$, $(2mn-1)\otimes m\otimes n\otimes 2$.
\end{corollary}

For $\h=\h_7\otimes \h_2\otimes \h_2\otimes
\h_2$, it follows from the above Corollary that it has a finite number of maximal equivalence class.
In contrast,  $\h$ has an infinite number of equivalence class \cite{VDMV02}.
Another notable case is $\h=\h_4\otimes\h_3\otimes \h_2$. We
know from \cite{MIY03} that $\h'=\h_2\otimes\h_3\otimes
\h_2$ has $8$ equivalence class. Thus by Theorem
\ref{finite}, $\h$ has at most $8$ different  maximal equivalence classes.
However, the exact number is strictly smaller than $8$ as
some equivalence classes do not correspond to any equivalence classes.
A careful investigation shows that $\h_4\otimes\h_3\otimes\h_2$ has
exactly $5$ maximal equivalence classes.

\section{Discussions and open problems}

We showed as our main result
that a multipartite quantum system is allowed to have a maximum entangled state
only when there is a subsystem whose dimension is no less than the total dimension of the rest
of the system. When this condition does not hold, there are multiple distinct maximal equivalence
classes. A complete classification of those
maximal states would be of great value, both theoretically and
practically. To this end, we provided a connection between the maximal
equivalence classes in a state space with the stochastic
equivalence classes in another state space of a smaller dimension. In
particular, we proved that when $d_1=d_2\cdots d_K-1$, there is
a one-to-one correspondence between the maximal equivalence classes of
$\h_{d_1}\otimes\cdots\otimes\h_{d_K}$ and the stochastic
equivalence classes of $\h_{d_2}\otimes\cdots\otimes\h_{d_K}$.
Various examples are studied to demonstrate the
applications of these results.

We conclude by proposing two directions for further investigations
that we consider of both theoretical and practical importance.
The first is to understand deeper the structure of
of partial orders on LOCC and equivalence class.
Structural results will not only deepen our understanding
of entanglement, but will also find applications for
establishing multipartite entanglement when there is no maximum state.

For example, which spaces have an infinite number of SLOCC equivalence
classes, or an infinite number of maximal classes?
For those spaces having a finite number of maximal equivalence classes,
the parties can share some number of each maximal states,
and use them later to generate arbitrary desired states.
Note that in this case the ratio of the output states and
the initial states will not be as efficient as the case when
a maximum state exists, unless the distribution of the output states
is known in advance. A second and related question is, given a
space that does not admit a maximum state, what is the ``smallest''
state outside the specified space yet is able
to generate an arbitrary state in that space? For instance, there
are two maximal equivalence classes in $2\otimes 2\otimes 3$, represented by the states
 $\ket{\Phi_1}$ and $\ket{\Phi_2}$ in Eqn.~(\ref{eqn:states223}).
Either state, however, can generate any state from $2\otimes 2\otimes 2$ through SLOCC.

A second direction is to consider approximate generation
of entangled states.
Are there spaces that do not have a maximum
state but have an ``approximate'' maximum state in the sense that
all other states can be approximated to an arbitrary small precision
through a LOCC protocol on that state?
Such an approximate state is as good as the precise state in practice.
Consider another setting where the parties wish to generate
a large number of a target state.
A solution is for them to share in bulk some initial state, since
many copies of a fixed state are likely to be cheaper to manufacture.
A natural question is, which initial state will offer the most efficient rate
of conversion in the worst case (over all possible target states)?
In particular, which spaces admit the best possible ratio of $1$ asymptotically?
Perhaps the notion of ``border rank'' (C.f. Chapter 15 in \cite{Burgisser}), the approximate version of tensor rank,
in algebraic complexity theory may be useful for tackling those intriguing problems.

\section{Acknowledgments}

We are grateful to Eric Chitambar and Zhengfeng Ji for helpful discussions and comments.

\bibliographystyle{abbrv}

\end{document}